\documentclass[preprint,12pt]{elsarticle}

\usepackage{graphicx}

\usepackage{amssymb}

\journal{Physica B}

\begin{document}

\begin{frontmatter}



\title{Thermoelectric figure of merit of $\tau$-type conductors of several donors}


\author{
H. Yoshino$^\mathrm{a}$, 
H. Aizawa$^\mathrm{b}$, 
K. Kuroki$^\mathrm{b}$, 
G.C. Anyfantis$^\mathrm{c}$, 
G.C. Papavassiliou$^\mathrm{c}$, 
K. Murata$^\mathrm{a}$
}

\address{
$^\mathrm{a}$Graduate School of Science, Osaka City University, Osaka 558-8585, Japan\\
$^\mathrm{b}$Department of Applied Physics and Chemistry, The University of Electro-Communications, Chofu, Tokyo 182-8585, Japan\\
$^\mathrm{c}$Theoretical and Physical Chemistry Institute, National Hellenic Research Foundation, Athens 11635, Greece
}

\begin{abstract}
Dimensionless thermoelectric figure of merit $ZT$ is investigated for two-dimensional organic conductors $\tau$-(EDO-{\it S,S}-DMEDT-TTF)$_2$(AuI$_2$)$_{1+y}$, $\tau$-(EDT-{\it S,S}-DMEDT-TTF)$_2$(AuI$_2$)$_{1+y}$ and $\tau$-(P-{\it S,S}-DMEDT-TTF)$_2$(AuI$_2$)$_{1+y}$ ($y \le 0.875$), respectively.  The $ZT$ values were estimated by measuring electrical resistivity, thermopower and thermal conductivity simultaneously.  The largest $ZT$ is 2.7 $\times$ 10$^{-2}$ at 155 K for $\tau$-(EDT-{\it S,S}-DMEDT-TTF)$_2$(AuI$_2$)$_{1+y}$, 1.5 $\times$ 10$^{-2}$ at 180 K for $\tau$-(EDO-{\it S,S}-DMEDT-TTF)$_2$(AuI$_2$)$_{1+y}$ and 5.4 $\times$ 10$^{-3}$ at 78 K for $\tau$-(P-{\it S,S}-DMEDT-TTF)$_2$(AuI$_2$)$_{1+y}$, respectively.  Substitution of the donor molecules fixing the counter anion revealed EDT-{\it S,S}-DMEDT-TTF is the best of the three donors to obtain larger $ZT$.
\end{abstract}


\end{frontmatter}


\section{Introduction}
\label{introduction}

Dimensionless thermoelectric figure of merit $ZT = S^2/\rho \kappa$ is known as a measure of efficiency of thermoelectric power generation, where $S$, $\rho$ and $\kappa$ denote thermopower (Seebeck coefficient), electrical resistivity and thermal conductivity, respectively.  Lower $\kappa$ is required to keep temperature difference between the ends of a thermoelectric material generating electricity by Seebeck effect.  Few studies have been reported for $\kappa$ of organic conductors from this view point, though organics generally have much lower $\kappa$ than conventional metals.

\begin{figure}[ht]
\begin{center}
\includegraphics[trim=20mm 100mm 20mm 30mm, scale=.25]{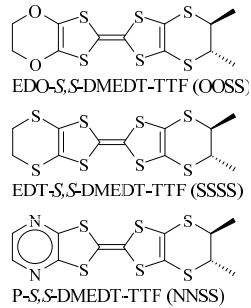}
\caption{Organic donors giving $\tau$-type organic conductors.}
\label{molecules}
\end{center}
\end{figure}

Recently some of the present authors have reported that $\tau$-(EDO-{\it S,S}-DMEDT-TTF)$_2$(AuBr$_2$)$_{1+y}$ ($y \leq 0.875$) shows relatively large $ZT$ of 0.042 at 130 K as an organic conductor \cite{yosh0901}, where EDO-{\it S,S}-DMEDT-TTF is an abbreviation of ethylenedioxy-{\it S,S}-dimethyl(ethylenedithio)tetrathiafulvalene (Fig.\ \ref{molecules}) and it is further shorten as OOSS below.  Among twelve donors giving $\tau$-type salts, those in Fig.\ \ref{molecules} are know to give salts of a common anion AuI$_2^-$.  It is worth comparing the three transport properties and $ZT$ of these AuI$_2$ salts with different donors to devise a strategy of synthesizing a donor giving more efficient thermoelectric organic materials.   Temperature dependence of $ZT$ of OOSS-AuI$_2$ and SSSS-AuI$_2$ (SSSS = EDT-{\it S,S}-DMEDT-TTF or ethylenedithio-{\it S,S}-dimethyl(ethylenedithio)tetrathiafulvalene) are reported elsewhere \cite{yosh1001, yosh1002}.  In this study $ZT$ of NNSS-AuI$_2$ (NNSS = P-{\it S,S}-DMEDT-TTF or pyrazino-{\it S,S}-dimethyl(ethylenedithio)tetrathiafulvalene) is determined and compared with the previous results.

\vspace{5mm}

\vspace{-5mm}
\section{Experimental}
\label{experimental}

Synthesis of NNSS was reported previously \cite{zamb9501}.  Crystals of NNSS-AuI$_2$ were grown by electrochemical oxidation method.  Samples studied are single crystals or polycrystals made of a few single crystals.  Sample sizes of OOSS-AuI$_2$, SSSS-AuI$_2$ and NNSS-AuI$_2$ are 0.78 $\times$ 0.34 $\times$ 0.02 mm$^3$, 1.22 $\times$ 0.36 $\times$ 0.02 mm$^3$ and 0.84 $\times$ 0.28 $\times$ 0.04 mm$^3$, respectively.

Crystal structures of the $\tau$-type conductors are made of alternating stacks of conducting and insulating layers.  The conducting layer contains 2:1 composition of donor molecules and anions, while the insulating layers does only anions corresponding to the composition $y$.  For OOSS-AuBr$_2$, elemental analysis suggests $y \sim 0.75$ \cite{papa9601}, while X-ray photograph revealed existence of a superstructure from which the upper limit of $y$ was estimated as 0.875 \cite{yosh0501}.  Crystal structure analysis is very difficult for all the $\tau$-type conductors due to the superstructure.

\begin{figure}[ht]
\begin{center}
\includegraphics[trim=5mm 110mm 5mm 15mm, scale=.35]{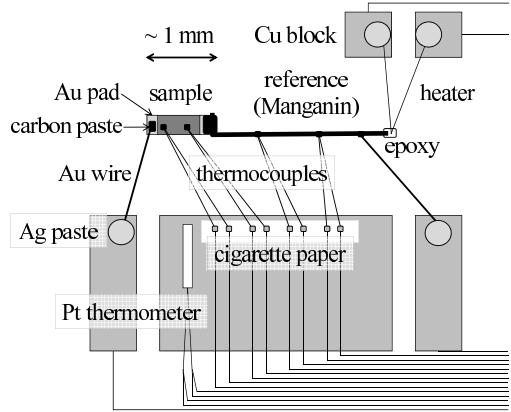}
\caption{Sample holder to measure electrical resistivity, thermopower and thermal conductivity simultaneously.}
\label{holder}
\end{center}
\end{figure}

Figure \ref{holder} shows a sample holder to measure $\rho$, $S$ and $\kappa$ simultaneously.   A conventional DC four-probe method was used to measure $\rho$.  Two of four pairs of Chromel-Constantan thermocouples (12.5 $\mu$m in diameter) were attached directly to the sample with carbon paste and the others were done to the Manganin wire (50 $\mu$m in diameter) connected to the sample in series to measure $\kappa$ by a steady-stated comparative method.  After making steady thermal gradient on the sample and Manganin wire by a heater, the temperature gradient on the sample and on the Manganin wire was measured by using each two pairs of the thermocouples.  Then thermoelectric power between the ends of the Chromel-sample-Chromel open circuit as well as between the ends of the Constantan-sample-Constantan one was also measured to determine $S$ of the sample by a so-called direct method.  These three transport coefficients were measured along two-dimensional conducting plane of the salts between 78 and 300 K at ambient pressure.  Details of the measurement technique are described elsewhere \cite{yosh0801}.

Thermal gradient to measure $\kappa$ and $S$ was typically between 0.15/$\l$ and 0.4/$\l$ K, where $l$ is the distance between the thermocouples on the sample and in the order of 0.5 mm.  The sample holder was covered by a copper radiation shield to minimize the heat loss by radiation.  The magnitude of $\kappa$ determined by the present method for Alumel is twice as large as that in literature \cite{sund9201} but shows almost the same temperature dependence between 78 and 300 K.  This implies that the latter gives most of error in the present measurement.

\section{Results and discussions}
\label{discussions}

\begin{figure}[ht]
\begin{center}
\includegraphics[trim=15mm 10mm 0mm 5mm, scale=.35]{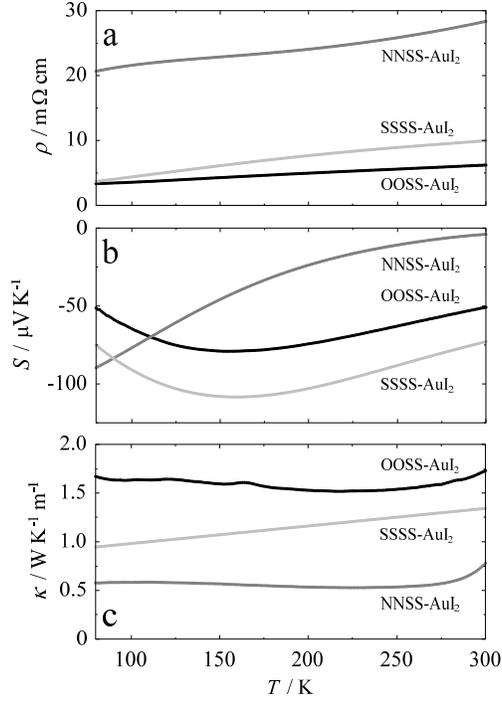}
\caption{Temperature dependence of (a) electrical resistivity, (b) thermopower and (c) thermal conductivity of $\tau$-(EDO-{\it S,S}-DMEDT-TTF)$_2$(AuI$_2$)$_{1+y}$ (OOSS-AuBr$_2$), $\tau$-(EDT-{\it S,S}-DMEDT-TTF)$_2$(AuI$_2$)$_{1+y}$ (SSSS-AuBr$_2$) and $\tau$-(P-{\it S,S}-DMEDT-TTF)$_2$(AuI$_2$)$_{1+y}$ (NNSS-AuBr$_2$), respectively.}
\label{rsk}
\end{center}
\end{figure}

Metallic $\rho$ was observed for all the materials, though the temperature dependence is very weak (Fig.\ \ref{rsk}(a)).  A recent band structure calculation on OOSS-AuBr$_2$ by Aizawa et al.\ shows that the band gap between the conduction band, where a small number of carriers exist, and the valence band is very small \cite{aiza1001}.  Then thermally activated carriers can make the temperature dependence of $\rho$ away from purely metallic.  In contrast, the metallic temperature dependence of the present salts down to 78 K suggests that the contribution of thermally excited carriers to electrical conduction is relatively small.

The sign of $S$ is negative as was observed for the other $\tau$-type conductors as in Fig.\ \ref{rsk}(b).  The room temperature value -4 $\mu$V K$^{-1}$ for NNSS-AuI$_2$ is the smallest among this class of materials, though $|S|$ reaches 90 $\mu$V K$^{-1}$ at 78 K.  OOSS-AuI$_2$ and SSSS-AuI$_2$ show the largest $|S|$ of 80 and 110 $\mu$V K$^{-1}$ at about 160 K, respectively.  Although $|S|$ of NNSS-Au$_2$ does not show maximum down to 78 K, it should turn to decrease below the temperature since $S$ becomes zero at 0 K.

The band structure calculation for OOSS-AuBr$_2$ suggests that there is a flat dispersion around the Fermi level \cite{aiza1001} and this is a common feature due to $\tau$-type donor packing.  Kuroki and Arita call this kind of band ``Pudding Mold''-type and pointed out that such an electronic system can give large $|S|$ in the order of 100 $\mu$V K$^{-1}$ as in Na$_x$CoO$_{2.9}$ \cite{kuro0701}.  Furthermore Aizawa et al. have succeeded to reproduce not only the magnitude but also the temperature dependence of $S$ of the $\tau$-type conductors by semi-classical calculation based on the band structure calculation \cite{aiza1001}.  It was also shown that temperature and magnitude of the maximum of $|S|$ are explained by competition between negative contribution by electrons in the conduction band and positive contribution by thermally excited carriers over the small band gap.

It should be noted that $|S|$ of NNSS-AuI$_2$ and OOSS-AuI$_2$ is much smaller than that of corresponding AuBr$_2$ and AuCl$_2$ salts.  In other words, a salt of a smaller anion gives larger $|S|$ \cite{yosh1001, yosh1002}.  Since $|S|$ of SSSS-AuI$_2$ is larger than that of OOSS-AuI$_2$, much larger $|S|$ than ever observed is expected for SSSS-AuBr$_2$ and SSSS-AuCl$_2$, though no one has ever reported these materials.

The accuracy of $\kappa$ is not so high as that of the electrical properties $\rho$ and $S$ due to possible thermal leak through the thermocouples as well as remaining gas in a vacuum chamber where the sample holder is contained.  It can be, however, concluded that $\kappa$ of these materials is in the order of 1 W K$^{-1}$ m$^{-1}$ as is close to that of the other organic conductors ever reported.  Contribution of electrical carriers to thermal conduction is estimated to be very small and in the order of \% when one uses $\rho$ for the estimation simply assuming Wiedemann-Franz law.  The small thickness ($\sim$0.02 mm) of the crystals, however, prevents us extracting more precise information by comparing the absolute values of $\kappa$ in Fig.\ \ref{rsk}(c).

\begin{figure}[t]
\begin{center}
\includegraphics[trim=5mm 125mm 5mm 20mm, scale=.35]{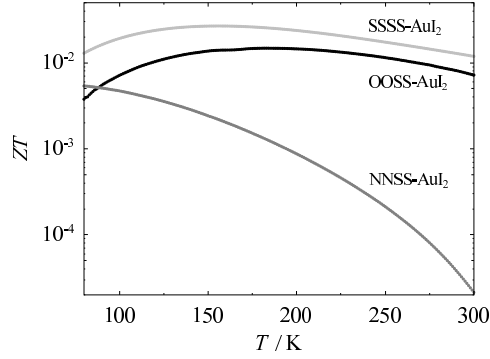}
\caption{Temperature dependence of dimensionless thermoelectric figure of merit of $\tau$-type conductors}
\label{zt}
\end{center}
\end{figure}

Figure \ref{zt} shows $ZT$ calculated from the data in Fig.\ \ref{rsk}.  The largest $ZT$ is 2.7 $\times$ 10$^{-2}$ at 155 K for SSSS-AuI$_2$, 1.5 $\times$ 10$^{-2}$ at 180 K for OOSS-AuI$_2$ and 5.4 $\times$ 10$^{-3}$ at 78 K for NNSS-AuI$_2$, respectively.  The small $S$ of NNSS-AuI$_2$ is the reason why its $ZT$ is much smaller than that of the others.  Substitution of oxygen atoms of OOSS with bigger sulfur ones gives SSSS and it should make overlap between neighboring molecular orbitals in the conduction layer of SSSS-AuI$_2$ larger than that in OOSS-AuI$_2$.  The observed larger $|S|$ of SSSS-AuI$_2$ than that of OOSS-AuI$_2$ is consistent with one of possible predictions that a wider ``Pudding Mold'' band results in larger $|S|$ as is opposite to that observed for conventional metals \cite{aiza1001} and also explains larger $ZT$ of SSSS-AuI$_2$ than that of OOSS-AuI$_2$.  Thus we can conclude that salts of donor molecules containing bigger heteroatoms like sulfur and, further, selenium with smaller anions like AuBr$_2^-$ and AuCl$_2^-$ are the best candidates to obtain much larger $ZT$ than 0.042 observed for OOSS-AuBr$_2$ at 130 K \cite{yosh0901}.

\section{Summary}
The dimensionless thermoelectric figure of merit $ZT$ was estimated by measuring $\rho$, $S$ and $\kappa$ of OOSS-AuI$_2$, SSSS-AuI$_2$ and NNSS-AuI$_2$, respectively.  It was found that NNSS-AuI$_2$ gives much smaller $ZT$ than that of the OOSS and SSSS salts due to much smaller $|S|$ at higher temperatures.  By comparing the results for the AuI$_2$ salts with different kinds of donors, it was found that SSSS is the best donor to get larger $ZT$ as well as larger $|S|$.  The ``Pudding Mold'' band model explains not only this tendency but also that a smaller anion gives larger $|S|$ as was reported before.  The present result can be utilized as a new guideline to develop novel thermoelectric materials based on organic molecules.

\section*{Acknowledgments}

This work was supported by Grant-in-Aid for Scientific Research on Priority Areas, gNanospaceh (Grant No. 19081011) from MEXT, Japan.




\begin{thebibliography}{00}


\bibitem{yosh0901}
H. Yoshino, G.C. Papavassiliou and K. Murata,
Synth. Met. (in press).

\bibitem{yosh1001}
H. Yoshino, A. Morimoto, G.C. Papavassiliou and K. Murata,
submitted to J. Phys. and Chem. Solids. 

\bibitem{yosh1002}
H. Yoshino, A. Morimoto, K. Ishida, G.C. Anyfantis, G.C. Papavassiliou, H. Aizawa, K. Kuroki and K. Murata,
submitted to J. Electron. Mater.

\bibitem{zamb9501}
J.S. Zambounis, J. Pfeiffer, G.C. Papavassiliou, D.J. Lagouvardos, A. Terzis, C.P. Raptopoulou, P. Delha\`es, L. Ducasse, N.A. Fortune and K. Murata,
Solid State Commun. {\bf 95} (1995) 211.

\bibitem{papa9601} G.C. Papavassiliou, D.J. Lagouvardos, J.S. Zambounis, A. Terzis, C.P. Rartopoulou, K. Murata, N. Shirakawa, L. Ducasse and P. Delhaes,
Mol. Cryst. Liq. Cryst. {\bf 285} (1996) 83.

\bibitem{yosh0501} H. Yoshino, K. Murata, T. Nakanishi, L. Li, E.S. Choi, D. Graf, J.S. Brooks, Y. Nogami and G.C. Papavassiliou,
J. Phys. Soc. Jpn. {\bf 74} (2005) 417.

\bibitem{yosh0801}
H. Yoshino, G.C. Papavassiliou and K. Murata,
J. Therm. Anal. Calorim. {\bf 92}  (2008) 457.

\bibitem{sund9201}
B. Sundqvist, 
J. Appl. Phys. {\bf 72} (1992) 539.

\bibitem{aiza1001}
H. Aizawa, K. Kuroki, H. Yoshino and K. Murata,
to be published in Physica B (this issue).

\bibitem{kuro0701}
K. Kuroki and R. Arita,
J. Phys. Soc. Jpn. {\bf 76} (2007) 083707.

\bibitem{bond6401}
A. Bond,
J. Phys. Chem. {\bf 68} (1964) 441.

\end{thebibliography}
\end{document}